\documentclass[manuscript]{aastex}
\usepackage{graphicx}
\usepackage{amsmath}
\usepackage[caption=false]{subfig}
\usepackage{natbib}
\usepackage{booktabs}

\begin{document}

\title{Prediction for IMAP: Revealing the Role of the Solar Magnetic Field in the Heliotail}

\author{M. Kornbleuth\altaffilmark{1,2}}
\affil{\altaffilmark{1}Astronomy Department, Boston University, Boston, MA 
02215, USA} 
\affil{\altaffilmark{2}Center for Space Physics, Boston University, Boston, MA 
02215, USA} 
\email{kmarc@bu.edu}

\author{M. Opher\altaffilmark{1,2}}

\begin{abstract}
The classic paradigm of the heliosphere considered the solar magnetic field as having a passive role in the heliosheath, yet \citet{Opher15} and \citet{Drake15} suggested that the solar magnetic field plays an active role by collimating the solar wind plasma. Previous work has suggested that high-latitude lobes observed by IBEX-Hi energetic neutral atom (ENA) maps are solely due to the fast/slow solar wind profile, yet other works have found that the solar magnetic field may play a role in organizing the lobes as well. This is the first study to show how the ENA maps at higher energies can disentangle the collimation from the solar cycle effects. Here, using an MHD solution with and without the solar magnetic field, we find that the fast/slow wind profile largely dictates the ENA profile up to $\sim$10-18 keV. Beyond this energy, the collimation effects are more prominent and the fast/slow solar wind profile no longer organizes the maps. The collimation creates two high latitude lobes. By 80 keV, our steady state heliosphere with collimation yields an enhancement of ENAs of the high-latitude tail lobes compared to the low-latitude region by a factor of $\gtrsim$ 1.3, whereas the uncollimated heliosphere yields a ratio $<$ 1. Therefore, IMAP should be able to identify whether the solar magnetic field collimates the solar wind in the deep heliotail.
\end{abstract}


\section{Introduction}
Since \citet{Parker61}, the classic paradigm for the heliosphere has been a heliosphere with a comet-like tail, extending for thousands of au. In this canonical picture, which was confirmed by other modeling efforts \citep{Baranov93, Pauls95, Zank96}, considered the solar magnetic field as having a passive role in the heliosheath. \citet{Opher15} challenged this perspective, and found that the solar magnetic field plays an active role in the heliosheath, collimating the solar wind plasma into two jets. This collimation was caused by the tension of the solar magnetic field confining the solar wind plasma and diverting the low-latitude solar wind to the north and south (see also \citet{Drake15}). While \citet{Opher15} suggested that the active role of the solar magnetic field yields a short, split-tail heliosphere, other models \citep{Pogorelov15,Izmodenov15} argue for a long-tail heliosphere. Regardless of the shape, all of these models found that the solar magnetic field collimates the plasma in the heliosheath. This was further confirmed by \citet{Kornbleuth21a}, that compared the split-tail and long-tail models of the heliosphere using identical boundary conditions. They found that the collimation of the heliosheath flows persisted regardless of the heliospheric tail.

Energetic neutral atom (ENA) modeling of the heliosphere is fundamental in investigating the global structure and processes of the outer heliosphere. To date, the IBEX-Hi instrument \citep{Funsten09} aboard IBEX \citep{McComas09a} and the INCA instrument aboard Cassini \citep{Krimigis09} have been the two primary ENA surveyors of the heliosheath. Combined, these two instruments have an energy range from approximately 0.52-55 keV, yet despite the large energy range they are limited in how deep they are able to probe into the heliotail. The depth is set by the distance where the parent ions originating from the supersonic solar wind are charge exchanged. At energies $<$55 keV, this length is $\sim$400 au, where the long and split tail heliospheres present identical solutions \citep{Kornbleuth23a}. To probe deeper than 400 au, one needs to use ENAs at energies $>$55 keV. For instance, various works \citep{Reisenfeld21,Kornbleuth21b,Kornbleuth23a} found that IBEX-Hi and INCA energies could not probe deep enough into the heliotail to identify the shape of the heliosphere. \citet{Kornbleuth23a} found that with ENAs at energies of 80 keV, distinguishing the shape of the heliosphere is possible. This relies on the assumption that the parent ion population is only accelerated at the termination shock. Ultimately, the enhanced energy range of the recently launched IMAP satellite \citep{McComas25}, with the IMAP-Ultra instrument capable of probing ENAs to 100 keV in energy, will shed significant light on the dynamics of the deep heliotail.

Early results from IBEX-Hi showed two high-latitude heliotail lobes and a deficit of flux at low latitudes manifesting as the port and starboard lobes in the heliotail starting at $\sim$ 2 keV \citep{McComas13}. Early explanations \citep{McComas13,Zirnstein16a,Zirnstein20} for this lobe structure suggested that it was the byproduct of the evolving fast/slow solar wind structure. \citet{Kornbleuth20} showed that, in conjunction with the fast/slow wind structure, the solar magnetic field also played an active role in organizing the high-latitude heliotail lobes in terms of their shape and intensity. \citet{Dayeh22} performed an investigation of the heliotail lobes, and found a cycling pattern correlating with changes in the solar cycle. \citet{Kornbleuth24} investigated the different sources for the cycling of the lobes, and found that the solar magnetic field was a critical component in this evolution.

IMAP will provide notable enhancements over the previous ENA observations in statistics and resolution. With the ability to probe to energies not previously observed, there is the question regarding the energy at which structures seen in the ENA flux in the tail is due to the solar wind structure versus solar magnetic field collimation of the solar wind. Here, we address this question by probing the effects of magnetic field collimation in high energy to identify the energies where the fast/slow solar wind structure no longer dictates the observed features. We also aim to identify potential ENA signatures to confirm whether the solar magnetic field is (or is not) playing a role in confining the solar wind in the heliosheath. In Section \ref{sec:models} we present the MHD models used within this work, notably focusing on a case with and a case without solar magnetic field. In Section \ref{sec:results}, we present the results of the comparison of the two MHD solutions via ENA modeling. In Section \ref{sec:Summary}, we discuss our results present a summary of our findings.

\section{Models} \label{sec:models}
In this section we review briefly the models used in this work. For ENA modeling, we use the model from \cite{Kornbleuth23b}. This ENA model has been shown to produce qualitative agreement with both the IBEX-Hi and INCA observations, with the exception of the heliotail ``Buckle'' in the 8.38 keV energy band of INCA that was suggested as being attributable to magnetic reconnection \citep{Kornbleuth26}. In order to isolate the effects of the solar wind profile and the solar wind collimation, we do not include the magnetic reconnection mechanism from \citet{Kornbleuth26} in our modeling here. For the MHD models, we use solutions from \citet{Kornbleuth21a}, which we briefly describe in Section \ref{sec:models}.

In this work, we use the BU kinetic-MHD model of the heliosphere \citep{Kornbleuth21a}. The BU kinetic-MHD model couples the Outer Heliosphere (OH) and Particle Tracker (PT) components from the Space Weather Modeling Framework (SWMF) \citep{Toth05}. It is a 3D kinetic-MHD model that solves for a single-ion plasma that combines the cold solar wind plasma and hot pick-up ions (PUIs) as one fluid. The neutral particles for the two models are solved kinetically using a Monte Carlo method \citep{Malama91,Izmodenov01,Tenishev21,Chen24}. Within this study, we primarily compare two different cases that utilize different versions of the BU model. The first case, described below, is the BU model from \citep{Kornbleuth21a}, that uses the AMPS particle tracking code to kinetically model neutrals \citep{Tenishev21,Michael21}. The other case, also described below, models neutrals kinetically using the particle-in-cell code FLEKS \citep{Chen23,Chen24}. Both kinetic neutral codes have been validated \citep{Chen24}, however the FLEKS code has the advantage of being able to perform future time-dependent studies.

For our two primary cases, use identical inner and outer boundary conditions from \citet{Izmodenov20}. The neutrals and ions originating in the interstellar medium (ISM) are assumed to have the same bulk velocity $v_{ISM}$ = 26.4 km/s (longitude = 75.4$^{\circ}$, latitude = -5.2$^{\circ}$ in ecliptic J2000 coordinate system) and temperature $T_{ISM}$ = 6530 K at the outer boundary, where the pristine ISM is not influenced by the heliosphere. The interstellar magnetic field intensity and orientation is assumed to be $B_{ISM}$ = 3.75 $\mu$G and $\alpha$ = 60$^{\circ}$, such that the magnetic field is aligned with the hydrogen deflection plane \citep{Lallement05} and $\alpha$ is the angle between the interstellar velocity and magnetic field vectors. The ISM proton density at the outer boundary is assumed to be $n_{p,ISM}$ = 0.04 cm$^{-3}$ and the neutral H atom density is $n_{H,ISM}$ = 0.14 cm$^{-3}$. While these boundary conditions differ from others in the literature (e.g., \citet{Zirnstein16b,Swaczyna24}, these density values were chosen by \citet{Izmodenov15} because they support agreement with observational diagnostics, such as the observed Voyager termination shock and heliopause asymmetries.

For the inner boundary conditions of our two primary cases, we use 22-year averaged solar cycle conditions (1995-2017). We take into account the Helio-latitudinal variations of the solar wind density and speed \citep{McComas00,Sokol13,Tokumaru21} , and the temperature is inferred by using the speed and an assumed sonic Mach number at Earth, of M=6.44 (corresponding to a solar wind temperature of $T_{SW}$ = 188,500 K) is used. While hourly-averaged solar wind data from the OMNI 2 dataset \citep{King05} is used for the density and speed in the ecliptic plane, the model also accounts for heliolatitudinal variations of the solar wind speed and density based on analysis of interplanetary scintillation (IPS) observations \citep{Tokumaru12} from 1990 to 2017. For heliolatitudinal variations of the solar wind mass flux, SOHO/SWAN full-sky maps of backscattered Lyman-alpha intensities are used \citep{Quemerais06, Lallement10,Katushkina13, Katushkina19}. Data from SOHO/SWAN are available from 1995 to the end of 2017, and an inversion procedure is used to obtain the solar wind mass flux as a function of time and heliolatitude. 

For the first case, we include the solar magnetic field based on the Parker solution with the radial component of the magnetic field strength being $B_{SW}$ = 37.5 $\mu$G at 1 AU. This is to reflect the ``realistic'' case where the solar magnetic field is present in the heliosphere. For the second case, we neglect the solar magnetic field to investigate how the absence of the field in the heliosheath will affect ENA maps, and by extension, how the solar wind profile itself will affect ENA maps. We also run a third case that is hydrodynamic (i.e. no solar or interstellar magnetic field), which we do not model with ENAs, but use it to further understand the MHD solution in the absence of the solar magnetic field. For our case with solar magnetic field, the field is treated as unipolar \citep{Opher15,Izmodenov15} to avoid spurious numerical dissipation of the solar magnetic field in the heliosheath. Other models, such as \citet{Pogorelov17} and \citet{Bera25}, include time-dependent solar wind variations that were shown to weaken or destroy the collimation, yet these works include reversals of the magnetic field that introduce large swaths of numerical dissipation at smaller distances. This was shown by \citet{Michael18} to affect the numerical solution. This work therefore sets the bounds for a prediction of the BU model by analyzing the extremes - a case of a strong solar magnetic field without dissipation and a case with no collimation present - to offer a road map for critical interpretations. If the models that have spurious numerical dissipation will to lead to the breakdown of collimation, then the results should trend to the uncollimated case as there would be an absence of collimation. Therefore, we emphasize that the results from this work are strictly a prediction from the BU model.

All cases use the same grid as described in \citet{Kornbleuth21a} for the BU model. However, for the case without solar magnetic field and the hydrodynamic case, we extend the 4.7 au grid resolution in the PT component to span from x=-280 au to 1500 au (the outer boundary) and the 4 au resolution in the OH component  to extend from -300 to 1500 au. We extend the grid in these cases to maintain a constant resolution for the longer heliotail. The heliospheric jets (or lobes) of the heliotail created by the collimation of the heliosheath plasma by the solar magnetic field each have a width of $\sim$200AU. We employ a second-order scheme in our MHD modeling, such that we have a numerical resolution of 2 grid cells corresponding to 8 AU; hence we have sufficient grid resolution to capture accurately the collimation.

\section{Results}\label{sec:results}

\begin{figure*}[t!]
\centering
  \includegraphics[scale=0.55]{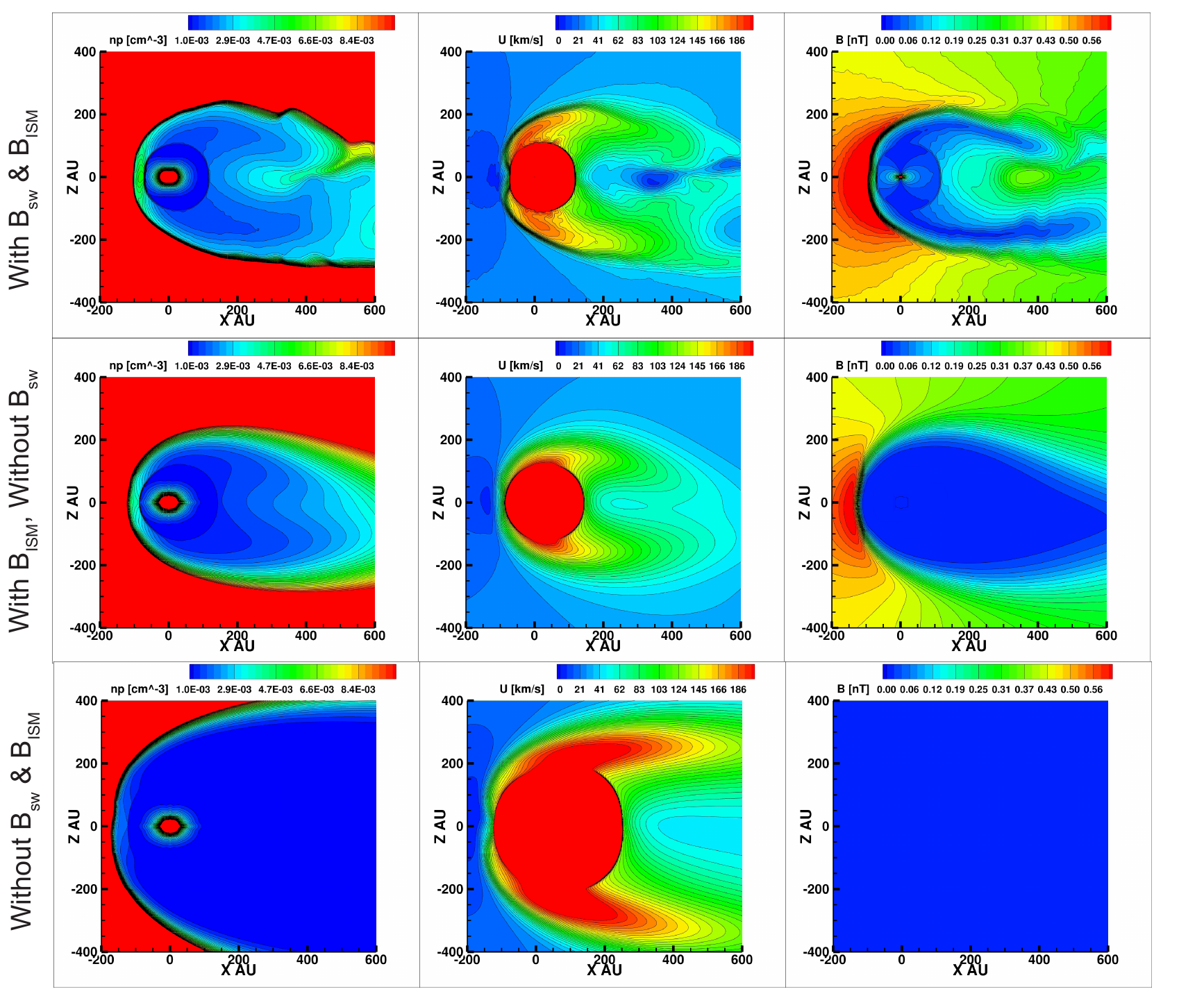}
  \caption{Collimation of the heliosheath flows by the solar magnetic field. We include the solution with solar magnetic field (top), without solar magnetic field (middle), and a hydrodynamic case (bottom) for plasma density (left), plasma speed (middle), and magnetic field intensity (right). The case with solar magnetic field demonstrates a collimated plasma and a split-tail heliosphere. The case without solar magnetic field displays no evidence of collimation and a long heliotail. The hydrodynamic case also shows no collimation, however it is not affected by the draping of the interstellar magnetic field and therefore the tail undergoes no compression effects unlike the case with solar magnetic field.}
  \label{fig:mhd}
\end{figure*}

\begin{figure*}[t!]
\centering
  \includegraphics[scale=0.60]{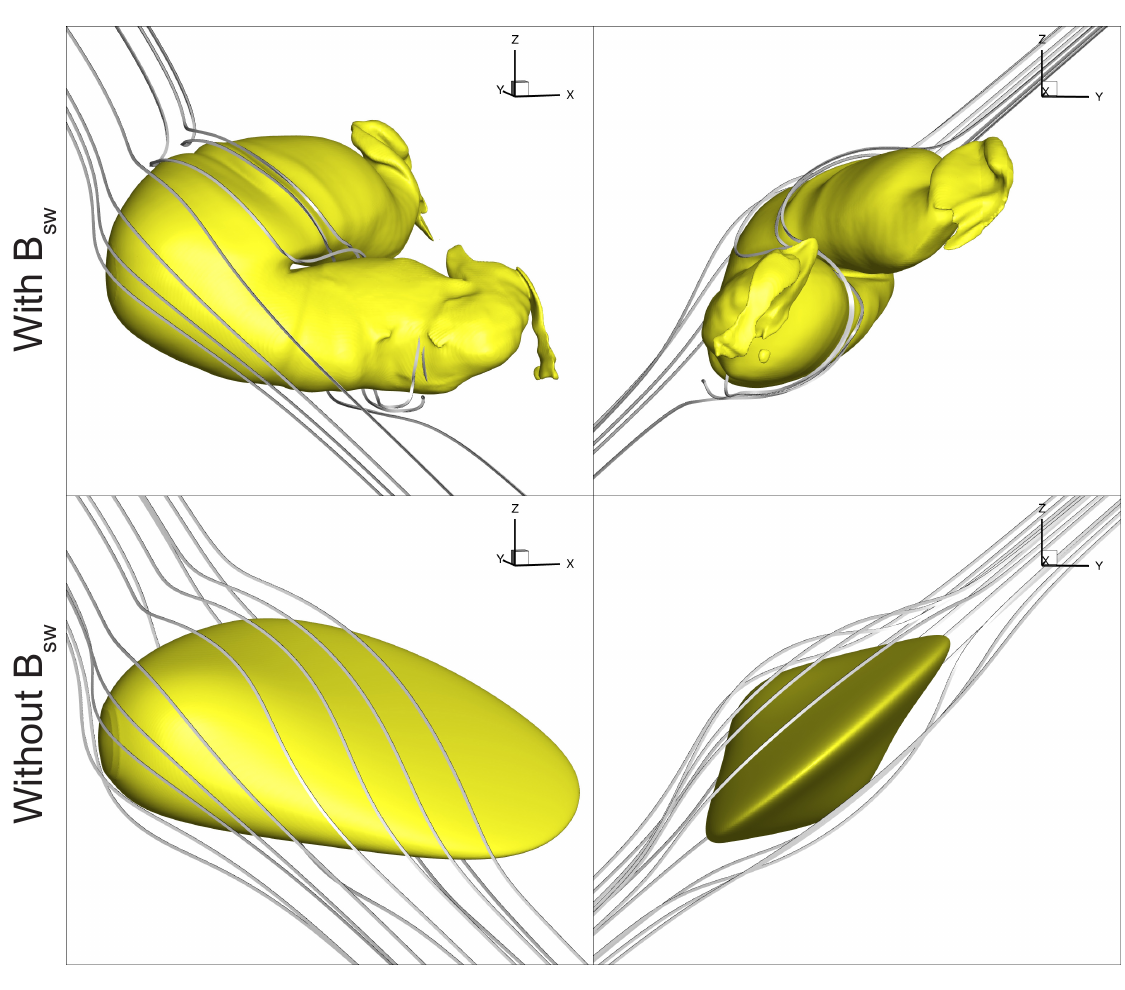}
  \caption{The draped interstellar magnetic field compresses the heliosphere. For the case with solar magnetic field (top), the magnetic pressure of the interstellar field draping along the heliopause is counterbalanced by the solar magnetic field. This is seen in the left and right panels showing the flank-view and the tail-view, respectively. For the case without solar magnetic field, the heliosphere experiences a thinning with distance due to the magnetic field draping. Ultimately, both cases orient along the orientation of the interstellar field, but the case without solar magnetic field exhibits a long-tail.}
  \label{fig:draping}
\end{figure*}

In Figure \ref{fig:mhd}, we show a comparison of the three cases described in Section \ref{sec:models}. We note that in the standard case with solar and interstellar magnetic field, a split-tail structure is seen, as previously identified in \citet{Opher15} and \citet{Kornbleuth21a}. In this model, the solar wind is collimated by the solar magnetic field into two jet-like structures streaming towards the north and the south. The interstellar flow and field bends the jets into two high latitude lobes. In contrast, the case without solar magnetic field does not demonstrate the same jet-like structure in the absence of collimation. While the fast solar wind fills the high-latitudes in the heliosheath and the slow solar wind fills the low-latitudes, with increasing distance the pressure from the draped interstellar field compresses the heliosphere. This effect eliminates the structure of the fast/slow wind in the deep heliotail. The draping can be seen in Figure \ref{fig:draping}, where the field lines overlay on the heliopause in both the cases with and without solar magnetic field. Yet in the presence of solar magnetic field with its magnetic tension, the heliosphere is able to resist the compression effects from the interstellar pressure. 

An interesting finding from the case without solar magnetic field is the shortening of the heliotail compared to the classic hydrodynamic case, also included in Figure \ref{fig:mhd}. While the hydrodynamic case, which is the classical perspective, yields a long tail extending beyond the outer boundary, we find that the heliotail in the case without solar magnetic field terminates around 700 au. As a point of comparison, the low-latitude heliopause in the split-tail model occurs at approximately 350 au. The difference in the tail lengths between the case without solar magnetic field and the hydrodynamic case is the magnetic pressure from the draped interstellar field, which is not present in the hydrodynamic case. While the meridional slice of the case without solar magnetic field seen in Figure \ref{fig:mhd} suggests a pinching of the tail, the 3D shape of the heliotail seen in Figure \ref{fig:draping} indicates that the heliotail has rotated to align with the interstellar magnetic field orientation.

\begin{figure*}[t!]
\centering
  \includegraphics[scale=0.43]{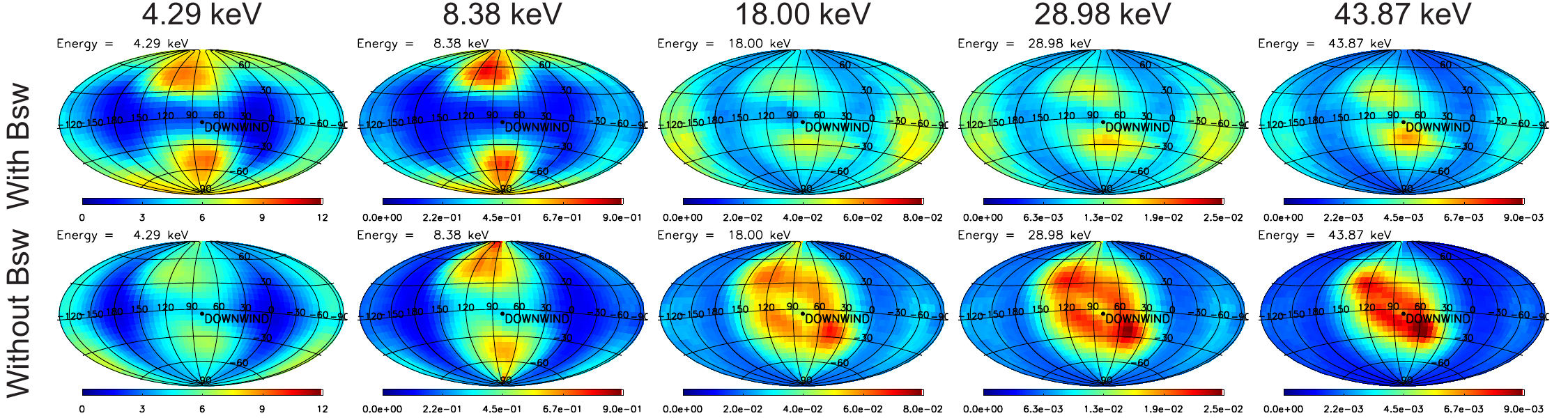}
  \caption{Collimation regulates ENA flux at high ENA energies. The ENA sky maps of flux are centered on the downwind (tail) direction in units of (cm$^{2}$ s sr keV)$^{-1}$. The top row reflects the case with solar magnetic field, while the bottom row reflects the case without solar magnetic field. While the high-latitude lobes in the 4.29 and 8.38 keV energy bands are weaker in the case without collimation, due to the presence of fast/slow wind the two cases are comparable. At higher energies, the fast/slow solar wind no longer regulates the ENA maps, and the lack of collimation in the case without solar magnetic field leads to the lobes ultimately converging into a central lobe by 43.87 keV.}
  \label{fig:maps}
\end{figure*}

In modeling the ENAs from the cases with and without solar magnetic field, we find notable differences between the results. In Figure \ref{fig:maps}, we present 5 different energy bands correlating with the highest energy of IBEX-Hi and the energy bands from INCA. We highlight these energies in particular to identify the energies at which the fast solar wind no longer regulates the observed structures seen in the ENA maps. By removing the solar magnetic field in the second case, we are able to probe directly what is attributed to the solar wind structure versus collimation effects when compared to the case with solar magnetic field. We find that in the 4.29 and 8.38 keV energy bands, the two cases present similar ENA maps in terms qualitatively and quantitatively. The primary difference between the two cases at these energies is the flux in the high latitude lobes is weaker in the case without solar magnetic field. This is due to the lack of collimation as previously noted in \citet{Kornbleuth20}. 

At the 18.00 keV energy band and above, we note a shift in the comparison between the two models. While the case with solar magnetic field maintains a two-lobe structure in the heliotail for all energies, the case without solar magnetic field shows a progression of the lobes towards lower latitudes. For the uncollimated case, this progression yields a convergence of the lobes into a single central lobe feature. We note that this central lobe is different than the one observed by IBEX \citep{Dayeh22}, as this occurs at significantly higher energies and relates to the compression of the heliosheath at far distances in the heliotail coupled with the lack of solar wind collimation. 

While a notable difference between the two cases is the shape of the heliotail, the merging of the lobes in the case without solar magnetic field cannot be attributed to the long-tail structure. As noted previously in \citet{Kornbleuth23a}, the Moscow model that has a long, comet-like tail and was compared directly to the BU model with a split-tail showed similar lobe structures up to energies of 43.87 keV. When collimation is present in both a short and long tail heliosphere, the maps are qualitatively and quantitatively similar. This is because the extinction prevents the observer at these energies from observing sufficiently far where differences between the two heliotail solutions can manifest. At the 18.00 keV energy band and above, the transmitted and reflected PUIs originating at the termination shock \citep{Zank10} cease to regulate the ENA production. Instead PUIs further accelerated at the termination shock by processes such as diffusive shock acceleration (DSA) \citep{Wang23,Kornbleuth23b} regulate the high-energy portion of the ion distribution in the heliosheath \citep{Giacalone21}.

To further illustrate this point, in Table \ref{tab:ratios}, we present ratios comparing the ENA flux from the northern heliotail lobe to the ENA flux at the downwind location. We use the northern lobe in the case with solar magnetic field as the reference location for the case without solar magnetic field. We find that in the 4.29 and 8.38 keV energy bands that the ratio between the flux in the lobes and downwind direction peak, reflecting the influence of the fast solar wind. However, at greater energies when the ion distribution is dominated by the DSA ions rather than the ions reflecting the fast/slow wind structure, the ratio between the lobes and downwind drop for both cases. The ratio remains $>1$ for the case with solar magnetic field, indicating the persistence of enhanced lobes at high latitudes. For the case without solar magnetic field, the ratio drops and becomes $<1$ by 43.87 keV when the lobes have converged. At 80 keV, the split-tail heliosphere yields an enhancement of ENAs in high-latitude tail lobes compared to the low-latitude region by a factor of 1.95, while the long-tail without solar magnetic field shows a roughly equal flux at low and high-latitudes in the tail region with a ratio of 0.90. In comparison, \citet{Kornbleuth23a} found the long-tail Moscow MHD solution with solar magnetic field included to have a ratio of 1.31. This result shows that while an ENA profile demonstrating enhanced flux at high-latitudes as compared to low-latitudes in the tail would indicate the presence of collimation in the heliosheath, this ratio being greater than unity would not necessarily indicate the shape of the heliotail. We note that these quantitative results are influenced by the assumptions within our model, such that it only considers energization of PUIs at the termination shock and uses a steady state solar cycle averaged profile of the solar wind. While this is not necessarily reflective of the realistic heliosphere, the quantitative aspect of our study does provide a baseline comparison of how collimation can affect ENA profiles.

\begin{figure*}[t!]
\centering
  \includegraphics[scale=0.45]{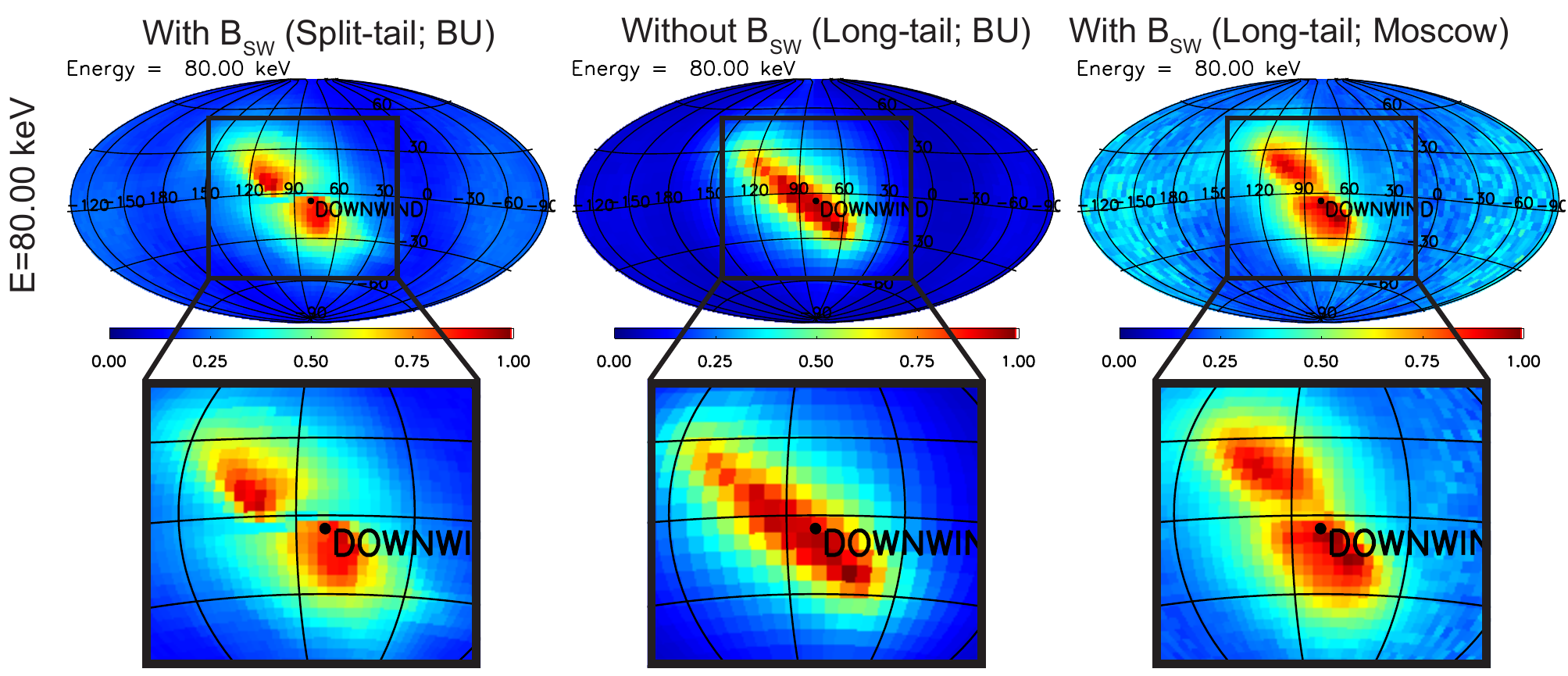}
  \caption{By 80 keV, the ENAs are entirely influenced by the collimation (or lack thereof), yielding two concentrated high-latitude lobes of enhanced ENA flux in the case with solar magnetic field, and a concentrated low-latitude central lobe in the case without solar magnetic field. The simulated ENA sky maps of flux using the BU MHD model are centered on the downwind (tail) direction and are normalized to the maximum flux in each map for the case with solar magnetic field (left) and without solar magnetic field (middle) at the 80 keV energy band. We also include the Moscow MHD, long-tail heliotail maps with solar magnetic field from \citet{Kornbleuth23a} in the right panels. In the bottom panels, we include a zoomed in view of the heliotail ENAs.}
  \label{fig:imapflux}
\end{figure*}

By 80 keV, we find that whereas two distinct lobes remain in the case with solar magnetic field, in the case without solar magnetic field the flux peaks in low latitudes as opposed to high latitudes (Figure 4). This is due to the increased depth into the heliosheath ($\sim$600 au) that can be observed \citep{Kornbleuth23a}. This is the energy in which the shape of the heliosphere can potentially be distinguished \citep{Kornbleuth23a}. Additionally, by this energy the ENA profile reflects the plane of the interstellar field and flow. For comparison, the previous investigation using the long-tail Moscow model, which can be found as the right panel of Figure \ref{fig:imapflux}, found a filled tail of ENAs at 80 keV. This was in contrast with the distinct split of the BU model with solar magnetic field, but the flux peaked as two-high latitude lobes as opposed to the central lobe seen in the case without solar magnetic field. 

\section{Discussion}\label{sec:Summary}
Our results indicate that by 18 keV, the fast/slow solar wind does not have a significant effect on the structuring of observed ENA profiles. By this energy, the collimation of the solar wind in the heliosheath by the solar magnetic field is the primary source of the ENA profile as two high latitude lobes in the tail direction. In contrast, without the solar magnetic field collimation, the ENAs with increasing energy should reveal the convergence of the two high latitude lobes into a central lobe feature. 

We note that while we identify the 18 keV energy band as the first energy that effects from collimation are more prominent than effects from the fast/slow wind profile in organizing ENA maps, there is uncertainty in this estimation. As the transition energy largely relies upon the velocity distribution of the fast solar wind considering its energy relative to the slow solar wind ions, time-dependent solar wind variations would have a notable effect in causing the transition energy to evolve with solar cycle. Additionally, inherent model assumptions such as the ion velocity distribution functions and the different assumed ion populations both tracing back to the termination shock or being created in the heliosheath can affect the transition energy. The flows in the heliosheath are also dictated by the assumed inner and outer boundary conditions of the MHD model, which can potentially affect the transition energy as well. Lastly, considering the modeling performed here correlated with the energy bands from IBEX and INCA missions, we do not explicitly identify the exact transition energy. We can only identify that the transition occurs between the 8.38 keV and 18.00 keV energy bands in terms of the governing effects. Considering the implications from our assumptions that can affect the transition, we emphasize that the existence of a transition energy is more relevant to this study than an exact predicted transition energy. Therefore, while we find the transition to occur in the range of $\sim$10-18 keV, this may be variable -- though we do not expect the transition to exceed 18 keV based on typical heliosheath properties.

The velocity distribution function of ions included in the ENA modeling can have an effect on the results of this work. Previous works \citep{Zirnstein18, Zirnstein25, Baliukin25} have presented modeling of ENAs using the Parker transport equation and modeling the velocity diffusion of PUIs in the heliosheath. Within our ENA model, we do not include velocity diffusion effects and are only considering PUI acceleration at the termination shock by using a velocity distribution function derived from hybrid modeling at the termination shock that was able to provide agreement with Voyager LECP observations \citep{Giacalone21}. Another recent work \citep{Fraternale26} proposed that the core thermal component of the solar wind plasma is hotter than measured by Voyager PLS observations downstream the termination shock, which can have important implications for re-evaluating the IBEX-Lo and IBEX-Hi energies (as well as IMAP-Lo/IMAP-Hi). This result however does not affect our current investigation as, based on our previous modeling efforts \citep{Wang23}, by the energies we are considering the parent ion population that largely regulates the ENAs is not the transmitted or reflected PUIs. Rather, a population further accelerated by DSA, which creates a power law and does not have a temperature dependence, largely regulates the ENAs at the energies being investigated here. Regardless, we emphasize that the effects of velocity diffusion and the underlying global heliosphere model can have implications on the velocity distribution function, which can influence the results. Therefore, this work serves as a template for other models that include acceleration in the heliosheath or a hotter core thermal population to investigate the disentanglement of the solar wind profile and the collimation at IMAP-Ultra energies.

For this work we focused only on time-independent boundary conditions while ignoring time-dependent solar wind effects. Time-dependent effects will have implications on this study both in terms of the solar wind profile and the solar magnetic field intensity. While the solar wind profile would affect the energy range of the transition to the ENA profile being primarily regulated by the magnetic field collimation as previously discussed, the varying intensity of the solar magnetic field would have a direct implication on the strength of the collimation. This would change the shape and intensity of the lobes, and could affect the ratio of flux in the heliotail lobes as compared to the low-latitude tail flux. Despite these potential effects, this study serves as an initial proof of concept to demonstrate that there is a limit to which the solar wind profiles organizes ENA maps. At sufficiently high energies, the collimation effects of the heliosheath plasma by the solar magnetic field can be observed via ENAs if present. In a future work, we will investigate the effects of a time-dependent solar wind and solar magnetic field on ENAs.

These results are significant for IMAP, as it builds up previous instruments to have an enhanced ability to observe ENAs up to approximately 100 keV. INCA observations previously indicated the persistence of high-latitude heliotail lobes up to 55 keV \citep{Dialynas13, Dialynas17, Sokol25}. With the improved statistics and extended energy range, the heliotail features seen by IMAP will be important in revealing whether the solar magnetic field is playing a critical role in collimating the flow deep in the heliosheath as first suggested by \citet{Opher15} and \citet{Drake15}. While the modeled maps here do not include solar wind variations, as shown the solar wind variations would only have effects up to a limited energy range, and by 18 keV the variations would have not regulate the ENA profile. Should IMAP see two distinct, separated lobes at high-energies ($\sim$80 keV), this would indicate that the solar magnetic field is collimating the solar wind plasma and therefore active role in shaping the heliosphere. If at these high energies IMAP instead sees a central lobe feature in the tail, this would indicate that the solar magnetic field is not collimating the solar wind plasma, and therefore is playing a passive role in the heliosheath. 

\acknowledgments
The authors are supported by NASA (grant No. 18-DRIVE18\_2-0029), Our Heliospheric Shield, 80NSSC22M0164. For more information about this center, please visit https://shielddrivecenter.com. This research was also supported by the International Space Science Institute (ISSI) in Bern, through ISSI International Team project ‘Physical Processes and Drivers of Particle Acceleration in the Heliospheric Tail As Seen Through ENAs and Interstellar Lyman-alpha Absorption’ (ISSI Team project \#24-613). M.O. would like to thank as well the support of a fellowship from the John Simon Guggenheim Memorial Foundation.


\newpage

\begin{table}[t!]
\centering
\begin{tabular}{ccc}
\hline
Energy [keV] & With B$_{SW}$ & Without B$_{SW}$ \\
 & Split-tail & Long-tail \\
\hline
 4.29 & 3.27  &  1.45    \\
 8.38 & 4.83  &  1.98    \\
 18.00 & 1.53  &  1.20    \\
 28.98 & 1.43  &  1.08    \\
 43.87 & 1.21  &  0.95    \\
 80.00 & 1.95  &  0.90    \\
\hline
\end{tabular}
\caption{Ratios of ENA flux from the northern lobe to the downwind direction for cases with and without solar magnetic field for the BU MHD model.}
\label{tab:ratios}
\end{table}

\end{document}